\pdfoutput=1
\documentclass[%
 reprint,
superscriptaddress,
 amsmath,amssymb,
 aps,
prb,
floatfix,
]{revtex4-2}

\usepackage{graphicx}
\usepackage{dcolumn}
\usepackage{bm}
\usepackage{color}
\usepackage{ulem}

\begin{document}


\title{Auxiliary-field quantum Monte Carlo method with quantum selected\\ configuration interaction}

\newcommand{\affilA}{%
Center for Quantum Information and Quantum Biology,
Osaka University, 1-2 Machikaneyama, Toyonaka, Osaka 560-0043, Japan
}%
\newcommand{\affilB}{%
Graduate School of Engineering Science, Osaka University, 1-3 Machikaneyama, Toyonaka, Osaka 560-8531, Japan
}%
\newcommand{\affilC}{Technology Strategy Center, TOPPAN Digital Inc., 1-3-3 Suido, Bunkyo-ku, Tokyo 112-8531, Japan}

\author{Yuichiro~Yoshida}
\email{yoshida.yuichiro.qiqb@osaka-u.ac.jp}
\affiliation{\affilA}
\author{Luca~Erhart}
\affiliation{\affilA}
\author{Takuma~Murokoshi}
\affiliation{\affilA}
\author{Rika~Nakagawa}
\affiliation{\affilC}
\author{Chihiro~Mori}
\affiliation{\affilC}
\author{Takafumi~Miyanaga}
\affiliation{\affilA}
\author{Toshio~Mori}
\affiliation{\affilA}
\author{Wataru~Mizukami}
\email{mizukami.wataru.qiqb@osaka-u.ac.jp}
\affiliation{\affilA}%
\affiliation{\affilB}%

\date{\today}

\begin{abstract}
We propose using the wave function generated by the quantum selected configuration interaction (QSCI) method as the trial wave function in phaseless auxiliary-field quantum Monte Carlo (ph-AFQMC).
In the QSCI framework, electronic configurations are sampled from the quantum state realized on a quantum computer. These configurations serve as basis states for constructing an effective Hamiltonian, which is then diagonalized to obtain the corresponding eigenstate. 
Using this wave function, ph-AFQMC is performed to recover the dynamical electron correlation across the whole orbital space.
The use of the QSCI trial wave function is expected to improve the feasibility of the quantum-classical (QC) hybrid quantum Monte Carlo approach [Nature, \textbf{603}, 416 (2022)].
We call this integrated approach QC-QSCI-AFQMC, or QSCI-AFQMC for short.
This method is validated across several molecular systems. 
For H$_2$O and a linear H$_4$ chain, we achieved chemical accuracy in most investigations relative to full configuration interaction while utilizing superconducting quantum computers at Osaka University and RIKEN.
Additionally, the application of QSCI-AFQMC to the O--H bond dissociation in an organic molecule highlights the complementary synergy between capturing static correlation on quantum hardware and incorporating dynamical correlation via classical post-processing. 
For the N$_2$, when QSCI-AFQMC is executed with a noiseless simulator, it ranks among the most accurate methods compared to various multireference electronic structure theories.
Although the proposed method is demonstrated using small active spaces on current quantum devices, the concept is not limited to few-qubit problems.
The QSCI-AFQMC can compete with state-of-the-art classical computational techniques, particularly in larger active spaces, displaying considerable potential for resolving classically intractable problems in quantum chemistry.
\end{abstract}

\maketitle



\section{Introduction \label{sec:intro}}

The rapid advancements in quantum computing technologies present a transformative opportunity for quantum chemistry, an essential field that investigates the intricate electronic states of matter and is recognized as a ``killer application'' for quantum devices.
In the past decade, the application of quantum computers in quantum chemical calculations has advanced significantly, nearing the capabilities of state-of-the-art classical methods~\cite{Peruzzo2014variational,Kandala2017hardware,Huggins2022unbiasing,Kanno2023quantum,Robledo2024chemistry}.

A prominent achievement in this domain is the quantum--classical hybrid (QC) quantum Monte Carlo approach reported by Huggins et al~\cite{Huggins2022unbiasing}.
This innovative framework integrates quantum computation into quantum Monte Carlo methods, such as the phaseless auxiliary-field quantum Monte Carlo (ph-AFQMC) method~\cite{Zhang2003quantum}, which is gaining traction in quantum chemistry.
The ph-AFQMC method is known for its great potential for capturing electron correlation while maintaining high parallelization efficiency~\cite{Zoran2024toward,Yifei2024GPU,Lee2022twenty,Sukurma2025self}.
However, ph-AFQMC results depend on the quality of the trial wave function, and there are limitations to the flexibility of the wave function available in classical computing.
Huggins et al. addressed this challenge by utilizing a quantum state as the trial wave function within their QC-AFQMC framework.
This approach is robust against noise, enabling successful quantum chemical calculations for systems such as N$_2$ and diamond, using up to sixteen quantum bits (qubits).

Still, a major challenge remains: the calculation of the overlap between the trial wave function and the walkers utilized in the quantum Monte Carlo method exhibits exponential scaling with respect to the size of the system.
In the original QC-AFQMC methodology, the trial wave function on a quantum computer was approximated by the classical shadows technique~\cite{Huang2020predicting}.
Although improved algorithms have been explored~\cite{Wan2023matchgate, Huang2024evaluating}, there is an ongoing need for the development of practically scalable approaches to tackle the increasingly larger and more complicated chemical problems.

Another advancement in this field is the quantum selected configuration interaction (QSCI) method~\cite{Kanno2023quantum,Nakagawa2024ADAPT}.
The QSCI determines the eigenvalues (energies) and eigenstates by diagonalizing an effective Hamiltonian, specifically the subspace Hamiltonian formed by electronic configurations sampled via a quantum computer.
The quantum computer is used to sample the electronic configurations required for accurately representing the electronic eigenstates of a molecule. 
Subsequently, the effective Hamiltonian can be efficiently diagonalized using classical computational resources.
Due to the efficient way to use the shot budget and noise resilience, this method has significantly expanded the number of qubits employed in quantum chemical calculations~\cite{Barison2024quantum,Kaliakin2024accurate,Liepuoniute2024quantum,Shajan2024quantum,Kaliakin2025implicit}. 
In fact, up to seventy-seven qubits were used in demonstrations for N$_2$, [Fe$_2$-S$_2$], and [Fe$_4$-S$_4$] clusters~\cite{Robledo2024chemistry}.
This achievement marks the highest utilization of qubits employed for quantum chemical calculations to date, highlighting the promising potential of the QSCI methodology.

However, a fundamental limitation of QSCI is its lack of some electron correlation arising from the inevitable use of an active space.
The active space consists of a select subset of molecular orbitals located near the Fermi level and are deemed chemically significant.
For highly precise quantum chemical calculations, a considerable number of molecular orbitals need to be employed to get quantitative results, which directly correlate with the number of qubits required. 
For example, in the Jordan--Wigner (JW) mapping---a prevalent fermion-to-qubit mappings---the total number of qubits corresponds to the count of spin orbitals.
The use of such an extensive set of orbitals is resource-intensive, making the introduction of an active space a necessity, even with the advancements in fault-tolerant quantum computers (FTQCs). 
Notably, this approximation has been utilized to estimate the quantum computational resources required for FTQCs~\cite{Markus2017elucidating,Li2019electronic,Lee2021even,Guang2025fast,Goings2022reliably}.
Furthermore, the development of methodologies to recapture electron correlation across the whole orbital space remains a pivotal challenge within this domain~\cite{Erhart2024coupled,Scheurer2024tailored,Shajan2024quantum,Khinevich2025enhancing}.
Therefore, a critical issue within QSCI persists: it inevitably lacks the consideration of electron correlation from orbitals that lie outside the designated active space.

In this study, we propose a ph-AFQMC method that uses the QSCI wave function---namely, the eigenstate of the effective Hamiltonian spanned by the electronic configurations sampled on a quantum computer---as the trial wave function. 
The QSCI wave function, situated within the active space framework, adeptly captures static electron correlation based on the principle of quantum superposition. 
Using this wave function as a trial state, the ph-AFQMC method effectively recaptures the deficiency of dynamical electron correlation across the whole orbital space.
This integrated methodology is referred to as QC-QSCI-AFQMC, in short, QSCI-AFQMC (continuing we use this short name). This method is scalable in terms of quantum resource efficiency and high flexibility, thanks to the QSCI methodology.
To assess the validity of QSCI-AFQMC, we investigate the O-H bond dissociation in H$_2$O and 2-hydroxyethyl methacrylate (HEMA) and the hydrogen chain dissociation, utilizing real quantum computing resources accessed via cloud platforms. Additionally, we examine the bond dissociation of the nitrogen molecule (N$_2$) using a computational emulator.

\section{Computational methods \label{sec:methods}}

In this section, we elaborate on the two pivotal components of our approach: ph-AFQMC~\cite{Zhang2003quantum} and QSCI~\cite{Kanno2023quantum}.

\subsection{Review of phaseless auxiliary-field quantum Monte Carlo}

AFQMC is a method for computing the ground-state wave function $|\Psi_{\rm g}\rangle$ for a given system through the process of imaginary-time evolution:
\begin{align}
    |\Psi_{\rm g}\rangle 
    = \lim_{n \to \infty} \left(e^{-\Delta\tau \hat{H}}\right)^n |\Psi_0\rangle, \label{eq:imag_evol}
\end{align}
where $|\Psi_0\rangle$ is the initial wave function.
Here, $\tau$ and $\Delta \tau = \tau/n$ denote the total imaginary time and its time step, respectively, and $\hat{H}$ denotes the Hamiltonian of the system.
The first-principles electronic structure Hamiltonian is given by
\begin{align}
    \hat{H} = \sum_{p,q=1}^N h_{pq}\hat{a}_p^\dagger \hat{a}_q + \frac{1}{2} \sum_{p,q,r,s=1}^N g_{pqrs}\hat{a}_p^\dagger \hat{a}_r^\dagger \hat{a}_s \hat{a}_q, \label{eq:H}
\end{align}
where $\hat{a}_p^\dagger$ and $\hat{a}_q$ represent the creation and annihilation operators for the $p$th and $q$th spin orbitals, respectively, while $N$ denotes the total number of spin-orbitals present in the system.
The coefficients $h_{pq}$ and $g_{pqrs}$ are the one- and two-electron integrals, respectively.

A widely adopted technique for handling the two-electron integral tensor is through a low-rank decomposition method, such as the Cholesky decomposition:
\begin{align}
    g_{pqrs} \approx \sum_{\gamma=1}^{N_\gamma} L_{pq}^\gamma L_{rs}^\gamma, \label{eq:Cholesky}
\end{align}
where $N_\gamma$ indicates the rank of the decomposition.
By substituting Eq.~(\ref{eq:Cholesky}) into Eq.~(\ref{eq:H}), the two-body term can be expressed as the square of a one-body operator:
\begin{align}
    \hat{H} = \hat{v}_0 - \frac{1}{2} \sum_{\gamma=1}^{N_\gamma} \hat{v}_\gamma^2, \label{eq:H_part2}
\end{align}
where
\begin{align}
    \hat{v}_0 &= \sum_{p,q=1}^N \left(h_{pq} - \frac{1}{2}\sum_{\gamma=1}^{N_\gamma} \sum_{t=1}^N L_{pt}^\gamma L_{tq}^\gamma \right)\hat{a}_p^\dagger \hat{a}_q \\
    \hat{v}_\gamma &= i \sum_{p,q=1}^N L_{pq}^\gamma L_{rs}^\gamma \hat{a}_p^\dagger \hat{a}_q. 
\end{align}

Using the Hamiltonian in Eq.~(\ref{eq:H_part2}), the short-time propagator $e^{-\Delta\tau\hat{H}}$ in Eq.~(\ref{eq:imag_evol}) can be factorized via a Trotter decomposition:
\begin{align}
    e^{-\Delta\tau\hat{H}} \approx e^{-\Delta\tau\hat{v}_0/2} e^{\Delta\tau\sum \hat{v}_\gamma/2} e^{-\Delta\tau\hat{v}_0/2} + \mathcal{O}(\Delta \tau^3). \label{eq:trotter_evol}
\end{align}

In AFQMC, one applies the Hubbard--Stratonovich transformation to the Trotterized propagator, introducing a vector ${\bm x}$ of auxiliary fields:
\begin{align}
    e^{-\Delta\tau\hat{H}} \approx \int{\rm d} {\bm x} \,p({\bm x})\hat{B}({\bm x},\Delta\tau) + \mathcal{O}(\Delta \tau^2),
\end{align}
where $p({\bm x})$ follows a normal distribution, and $\hat{B}({\bm x}, \Delta\tau)$ represents an effective one-body propagator: 
\begin{align}
    \hat{B}({\bm x},\Delta\tau) = e^{-\Delta\tau\hat{v}_0/2} e^{\sqrt{\Delta\tau}\sum x_\gamma \hat{v}_\gamma} e^{-\Delta\tau\hat{v}_0/2}.
\end{align}
As $\hat{B}({\bm x}, \Delta\tau)$ depends on the auxiliary field ${\bm x}$, it can incorporate the electron correlation effects.
In practice, a force bias is added to ${\bm x}$ to improve both stability and precision.

The ground-state wave function at imaginary time $\tau$, $| \Psi (\tau) \rangle$, is represented as
\begin{align}
    | \Psi (\tau) \rangle = \sum_{i=1}^{N_w} w_i \frac{|\Phi_i' (\tau) \rangle}{\langle \Psi_{\rm T} | \Phi_i' (\tau)\rangle},
\end{align}
where $\{ |\Phi_i'(\tau)\rangle \}$ is a set of walkers, $w_i$ is the weight of the $i$th walker, and 
$N_w$ is the number of walkers.
The trial wave function is represented as $|\Psi_{\rm T}\rangle$.

In the framework of AFQMC, importance sampling is utilized to evolve each walker in imaginary time with respect to $|\Psi_{\rm T}\rangle$.
After propagating a walker over a time increment $\Delta\tau$, its state is evaluated based on the overlap ratio:
\begin{align}
    S_i =  \frac{\langle\Psi_{\rm T}|\hat{B}({\bm x}, \Delta\tau)|\Phi_i'(\tau)\rangle}{\langle\Psi_{\rm T}|\Phi_i'(\tau)\rangle}.
\end{align}
The magnitude of $S_i$ indicates the significance of the walker, while its phase conveys any complex contributions introduced by the propagator.
The cumulative effect of complex phases produces the phase problem, which can severely increase the statistical errors in the computed expectation values.
To mitigate this complication, ph-AFQMC retains only the real component of $S_i$ during the updates of the walker weights:
\begin{align}
    w_i(\tau+\Delta\tau) &\propto w_i(\tau)|S_i|\max\{0, \cos(\Delta\theta)\}, \label{eq:weight} \\
    \Delta\theta &= \arg\left({\rm Re}[S_i]\right).
\end{align}
Note that the notation ``$\propto$'' is employed here, as the force bias factor is also factored in updating the weights.
This is the phaseless approximation.

\subsection{Review of quantum selected configuration interaction}

\subsubsection{Quantum selected configuration interaction}

QSCI constructs an approximate eigenfunction $| \Psi_{\rm QSCI} \rangle$ by sampling bit strings from a quantum state on a quantum computer~\cite{Kanno2023quantum}.
The QSCI wave function is written as
\begin{align}
    | \Psi_{\rm QSCI} \rangle = \sum_{i=1}^R c_i | \Phi_i \rangle, \label{eq:QSCI}
\end{align}
where the bit strings are assumed to be sorted in descending order based on their sampling frequency and are truncated to retain only $R$ electron configurations.
Each term $| \Phi_i \rangle$ corresponds to the $i$th bit string, while $c_i$ denotes its expansion coefficient.

The coefficients $\{c_i\}$ are determined by solving an eigenvalue equation within the subspace spanned by the sampled bit strings:
\begin{align}
    \hat{H}^{{\rm eff}}\bm{c} = E\bm{c},
\end{align}
where the effective Hamiltonian $\hat{H}^{{\rm eff}}$ is built from the original Hamiltonian $\hat{H}$ by projecting onto the sampled configurations.
Specifically, 
\begin{align}
    \hat{H}_{ij}^{{\rm eff}} := \langle \Phi_i | \hat{H} | \Phi_j \rangle, \label{eq:Heff_2}
\end{align}
and $E$ is the eigenvalue obtained from the equation, representing the energy of the system.

When an active space is introduced, $\hat{H}$ is replaced with the corresponding active-space Hamiltonian. The QSCI solution then becomes part of the wave function spanning all orbital degrees of freedom:
\begin{align}
    |\Psi_{\rm QSCI}\rangle = |\Psi^{(\rm virt)}\rangle \otimes |\Psi_{\rm QSCI}^{(\rm active)}\rangle \otimes |\Psi^{(\rm core)}\rangle,
\end{align}
where the entire orbital space is systematically classified into three categories: core orbitals (occupied by electrons), active orbitals (active), and virtual (unoccupied) orbitals (virt). Under this approximation, the QSCI wave function in the active space, $|\Psi_{\rm QSCI}^{(\rm active)}\rangle$, is given by Eq.~(\ref{eq:QSCI}).

\subsubsection{Use of Cartesian product states}

A naive QSCI approach may suffer from insufficient sampling, which can lead to violations of spin symmetry in the resulting wave function. To expand the space that can be sampled efficiently, we adopt Cartesian product states of subsystems corresponding to each spin.
Specifically, we write the states as 
\begin{align}
    |\Phi_i \rangle &= |n_{1\alpha,i}n_{1\beta,i}n_{2\alpha,i}n_{2\beta,i}\cdots n_{N\alpha,i}n_{N\beta,i}\rangle \\
    &= |n_{1\alpha,i}n_{2\alpha,i}\cdots n_{N\alpha,i}\rangle|n_{1\beta,i}n_{2\beta,i}\cdots n_{N\beta,i}\rangle \\
    &=: |\Phi_i^{(\alpha)} \rangle |\Phi_i^{(\beta)} \rangle,
\end{align}
where $n_{k\sigma,i} \in \{0, 1\}$ denotes the occupation number of the spin orbital associated with the $k$th spatial orbital and spin $\sigma$ in the $i$th configuration. 
Although various mappings from fermionic states to qubit representations can be utilized in a quantum computing framework, all states are mapped to those in the JW mapping when diagonalizing the effective Hamiltonian.
Here, $|\Phi_i^{(\sigma)} \rangle$ refers to the subsystem associated with the spin $\sigma$.

When building the effective Hamiltonian outlined in Eq.~(\ref{eq:Heff_2}), we employ the Cartesian products of the spin subsystems:
\begin{align}
    \{|\tilde{\Phi}_s\rangle\} = \left\{ 
    \begin{matrix}|\Phi_1^{(\alpha)} \rangle \\
    |\Phi_2^{(\alpha)} \rangle \\
    \vdots \\
    |\Phi_j^{(\alpha)} \rangle \\
    \vdots
    \end{matrix}
    \right\} \times \left\{
    \begin{matrix}|\Phi_1^{(\beta)} \rangle \\
    |\Phi_2^{(\beta)} \rangle \\
    \vdots \\
    |\Phi_k^{(\beta)} \rangle \\
    \vdots
    \end{matrix}
    \right\} 
\end{align}
instead of utilizing the original states $\{|\Phi_i\rangle\}$. Finally, the effective Hamiltonian is constructed through the following expression: 
\begin{align}
    \mathcal{H}_{st}^{{\rm eff}} = \langle \tilde{\Phi}_s | \mathcal{H} | \tilde{\Phi}_t \rangle. \label{eq:Heff_3}
\end{align}

\section{Computational details} \label{sec:comput_details}

We employed PySCF v2.2.1~\cite{PySCF,PySCF2} for self-consistent field, coupled-cluster singles and doubles (CCSD), coupled-cluster singles, doubles, and perturbative triples (CCSD(T)), complete active space configuration interaction (CASCI), complete active space self-consistent field (CASSCF), full configuration interaction (FCI), and selected configuration interaction (SCI) calculations. 
The \texttt{select\_cutoff} threshold for the SCI calculation was set to 0.0005.
The cc-pVDZ basis set was used unless otherwise specified.
The ph-AFQMC calculations were performed using ipie v0.6.2~\cite{Malone2023ipie}.

To construct and diagonalize the effective Hamiltonian, we utilized the Cartesian product of configurations for each spin through the \texttt{kernel\_fixed\_space} function in PySCF. Previous studies have also used the same function in their implementations~\cite{Robledo2024chemistry,qiskit-addon-sqd}. Notably, this approach does not directly depend on sampled configurations, and its implications are discussed in Section~\ref{sec:H2O}.

Variational quantum eigensolver (VQE) calculations were simulated classically via chemqulacs~\cite{chemqulacs}.
The resulting optimized quantum circuits were subsequently executed on actual quantum devices for sampling.
The quantum computations for O--H dissociation were performed using the superconducting quantum computer developed at Osaka University (OU): this device is the same as used in a previous experimental report on selective-excitation-pulse~\cite{Matsuda2025selective}.
For the quantum computation of H$_4$, we used the superconducting quantum computer A, located at RIKEN.
The chip design for this hardware was detailed in the work of Spring et al.~\cite{Spring2024fast}.
Both quantum computers featured 64-qubit chips with an overall architecture described in an earlier work~\cite{tamate2022toward}.

For quantum computation tasks across various cloud-based quantum devices, we leveraged the capabilities provided by
Quri~Parts~\cite{quriparts} (v0.14.0, v0.19.0, and v0.20.0) and Quri~Parts~riqu~\cite{quripartsriqu} according to the particular hardware.
Unless otherwise noted, we adopted the symmetry-conserving Bravyi--Kitaev (SCBK) mapping~\cite{Bravyi2017tapering} for the fermion-to-qubit mapping and hardware efficient ansatz~\cite{Kandala2017hardware} for VQE.

\section{Results and discussion} \label{sec:results}

\subsection{O--H bond dissociation\label{sec:OH}}

To validate the effectiveness of our approach, we initially applied QSCI-AFQMC to the O--H single bond dissociation, which is a frequently used benchmark setting in quantum chemistry. Our previous investigations revealed that the ph-AFQMC method in classical computing could accurately describe O--H dissociation~\cite{Yoshida2025auxiliary}, making it suitable for showcasing our quantum-classical hybrid methodology.

\subsubsection{H\textsubscript{2}O\label{sec:H2O}}

\begin{figure}[ht]
    \centering
    \includegraphics[width=1.0\linewidth, bb=0 0 461 346]{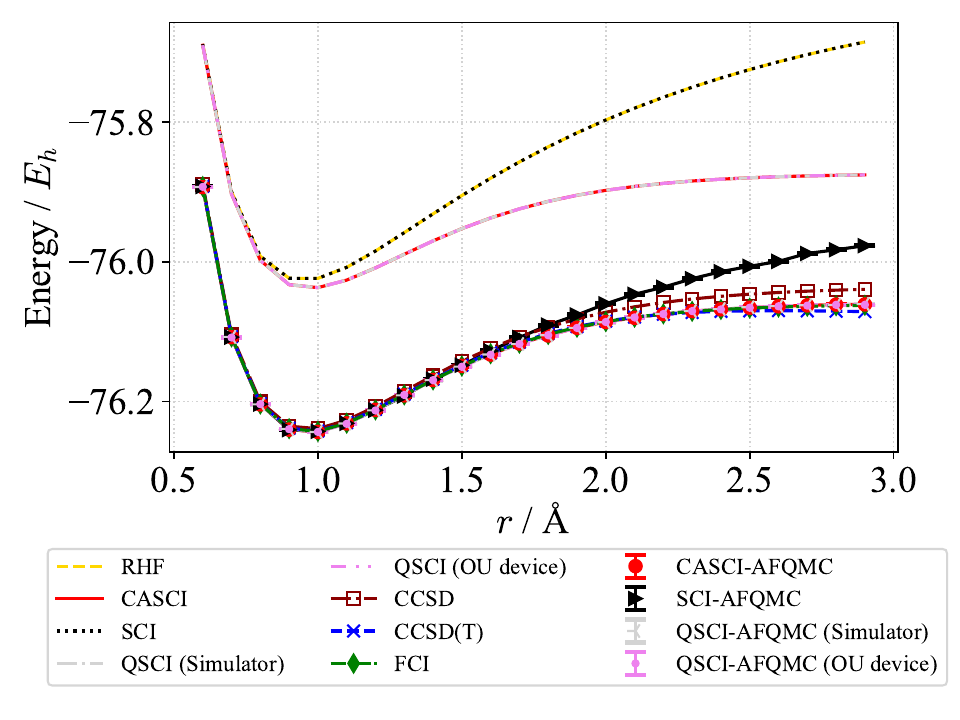}
    \caption{Energy curves for O--H dissociation in H$_2$O at the cc-pVDZ basis set. The active space is (2e, 2o). ``OU Device'' denotes the superconducting quantum computer located at Osaka University.}
    \label{fig:h2o}
\end{figure}

Figure~\ref{fig:h2o} shows the potential energy curves for O--H bond dissociation in H$_2$O, where we started from the stable molecular geometry optimized at MP2=FULL/cc-pVDZ level~\cite{CCCBDB} and varied the length of one O--H bond.
This figure illustrates how each method incorporates electron correlation.
First, the QSCI result matches CASCI, which refers to the exact solution in this (2e, 2o) active space. These calculations capture static correlation, which is not captured by restricted Hartree-Fock (RHF), and describe the O--H bond dissociation curve reasonably.
In contrast, a conventional SCI calculation (without quantum sampling) reproduces the RHF result, indicating that SCI selects only the HF configuration.

Figure~\ref{fig:h2o} further demonstrates that methods such as ph-AFQMC, CCSD, CCSD(T), and FCI yield lower energy values.
In the ph-AFQMC calculations, CASCI, SCI, and QSCI wave functions were used as trial wave functions, all of which produced energy curves significantly lower than those of the standalone methods.
This improvement reflects the dynamical correlation recovered from the space outside the active orbitals.
Importantly, QSCI-AFQMC closely mirrors the results obtained from both CASCI-AFQMC and FCI. Although CCSD and CCSD(T) yield comparable results, they differ from QSCI-AFQMC in the dissociation region, where deviations may occur. Notably, SCI-AFQMC exhibits larger errors at longer bond distances because the SCI curve aligns with that of RHF, a single-reference method.

\begin{figure}[ht]
    \centering
    \includegraphics[width=1.0\linewidth, bb=0 0 461 346]{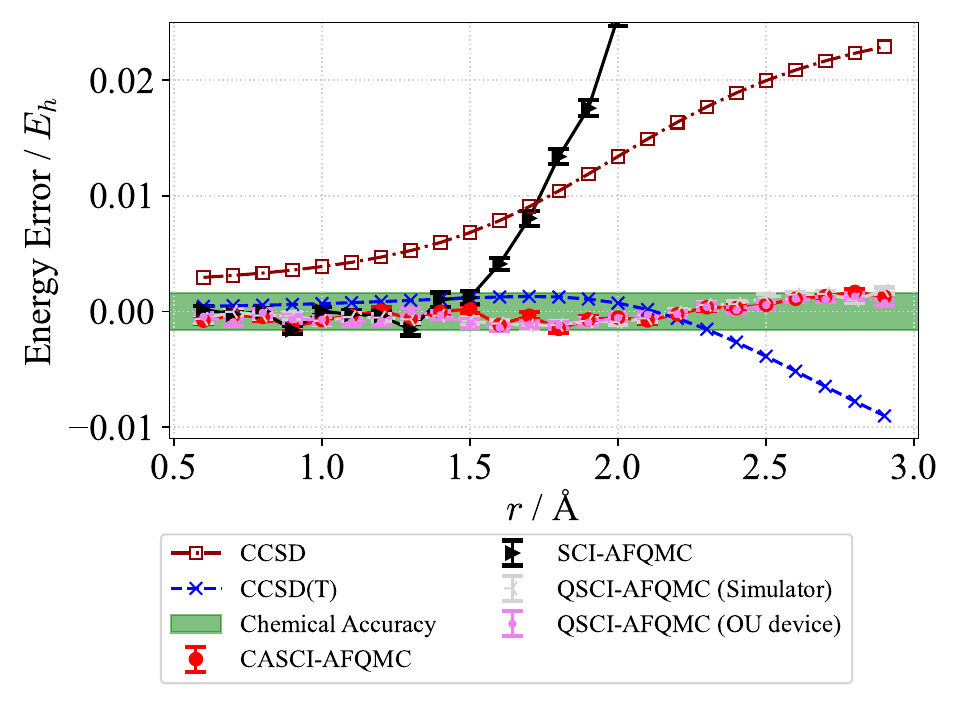}
    \caption{Energy error relative to FCI for O--H dissociation in H$_2$O. Calculations are at the cc-pVDZ level with an active space of (2e, 2o). ``OU Device'' denotes the superconducting quantum computer at Osaka University.}
    \label{fig:h2o_error}
\end{figure}

For a more detailed view of the accuracy, Figure~\ref{fig:h2o_error} illustrates the energy deviations from FCI.
Both simulated and device-based QSCI-AFQMC methods typically deviate from FCI by less than the chemical accuracy threshold of 1~kcal/mol $\sim 1.6 \times 10^{-3} E_h$ across most bond distances, matching the performance of CASCI-AFQMC.
In contrast, the error of CCSD exceeded the chemical accuracy for all points, and CCSD(T) failed at larger O--H bond distances.
As discussed earlier, the SCI-AFQMC method exhibits significant deviations in the dissociation region, where multireference effects become increasingly influential.

\begin{figure}[ht]
    \centering
    \includegraphics[width=1.0\linewidth, bb=0 0 576 360]{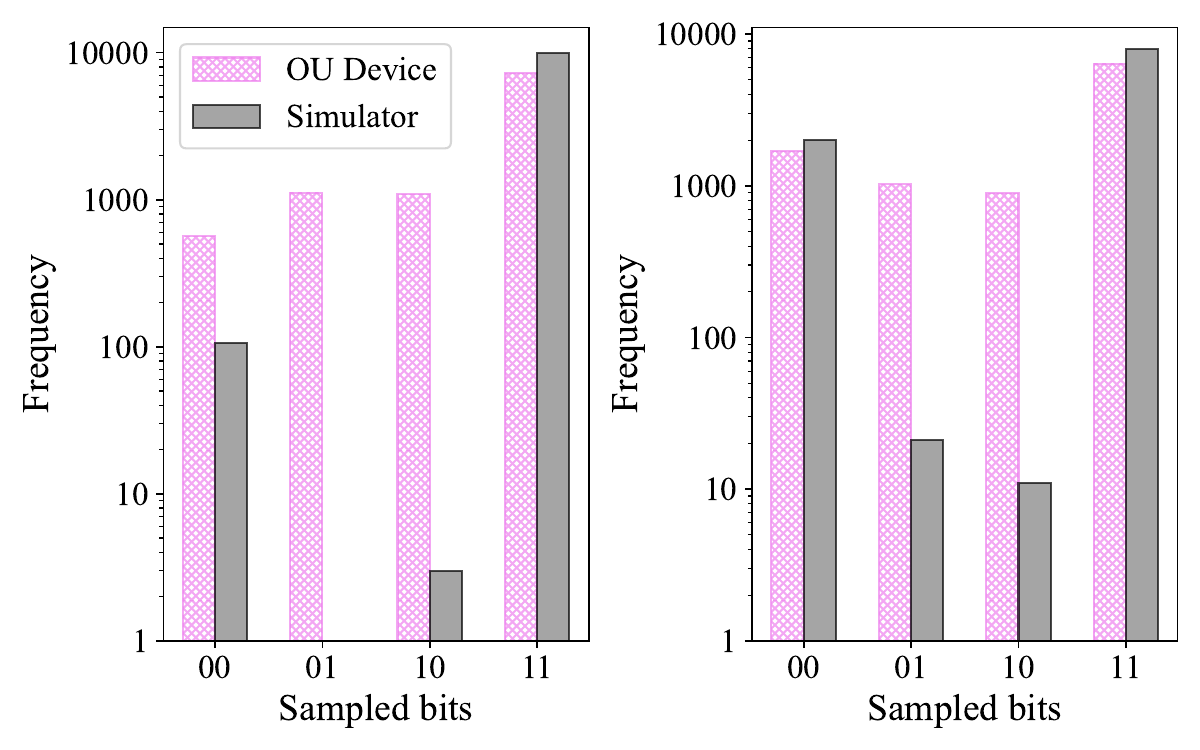}
    \caption{Sampling results for H$_2$O at $r=1.0$ {\rm \AA} (left) and $r=2.0$ {\rm \AA} (right). ``OU Device'' denotes the superconducting quantum computer at Osaka University.}
    \label{fig:h2o_sample}
\end{figure}

To delve deeper into the details of the QSCI, we present the bit sampling results obtained from the quantum computer (Figure~\ref{fig:h2o_sample}).
Quantum computations were conducted at each bond distance $r$ shown in Figures~\ref{fig:h2o} and~\ref{fig:h2o_error}, and here we highlight the cases for $r=1.0$ {\rm \AA} and $r=2.0$ {\rm \AA}.
For these calculations, we utilized the SCBK mapping, which preserves both particle number and spin symmetries, thereby reducing the required qubits to a mere two.
The following shows how the two-qubit strings map to the strings of the occupation-number basis:
\begin{align}
\begin{aligned}
    00 &\rightarrow 1100 \\
    10 &\rightarrow 0110 \\
    01 &\rightarrow 1001 \\
    11 &\rightarrow 0011. \label{eq:scbk2e2o}
\end{aligned}
\end{align}

Due to hardware noise, the device sampling results appear more randomized compared to those generated by the noiseless simulator.
Qualitatively, the HF configuration $| 0011 \rangle$ emerges as the most frequently observed bitstring in both cases. However, quantitative discrepancies in the sampling outcomes are attributed to the effects of noise.
We included all sampled configurations and constructed Cartesian product states.
As the size of the active spaces increases, the exponential growth in the number of configurations poses substantial challenges for effective sampling.
Enhancing gate and readout fidelity, alongside the implementation of more efficient QSCI strategies~\cite{Nakagawa2024ADAPT,Sugisaki2024Hamiltonian,Mikkelsen2024Quantum}, will be critical in advancing future research efforts.

\subsubsection{HEMA}

\begin{figure}[ht]
    \centering
    \includegraphics[width=1.0\linewidth, bb=0 0 851 341]{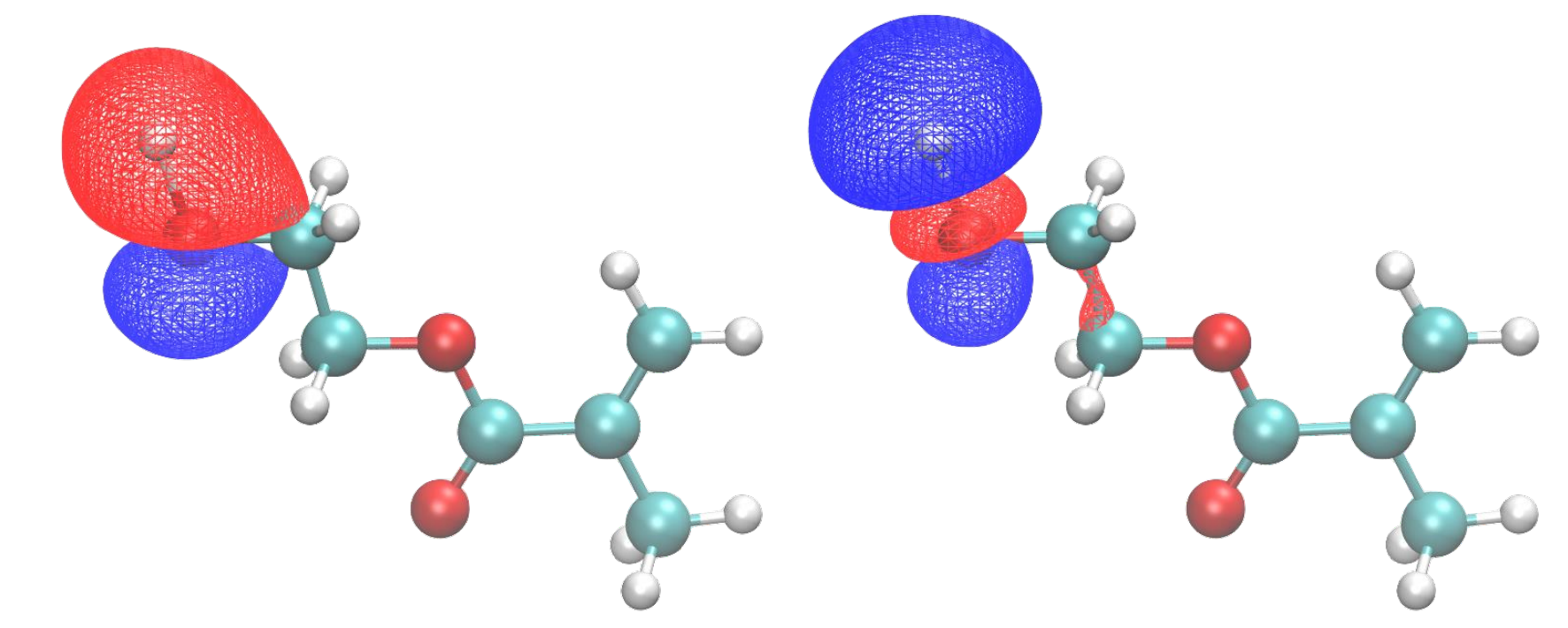}
    \caption{Molecular structure at $r=1.4$\,\AA\, and active space orbitals of HEMA, rendered by VMD~\cite{humphrey1996vmd}.}
    \label{fig:hema_mo}
\end{figure}

To illustrate that the proposed method can handle a large number of molecular orbitals while restricting the active space, we considered HEMA, an organic compound widely used in polymer chemistry. 
The molecular geometry is the same as that used in our previous study~\cite{Yoshida2025auxiliary}, varying the O--H bond distance $r$.
For the O--H bond, the $\sigma/\sigma^*$ molecular orbitals, as illustrated in Figure~\ref{fig:hema_mo}, were selected through the atomic valence active space method~\cite{Sayfutyarova2017automated}, whereas the total number of molecular orbitals for ph-AFQMC was 222.

\begin{figure}[ht]
    \centering
    \includegraphics[width=1.0\linewidth, bb=0 0 461 346]{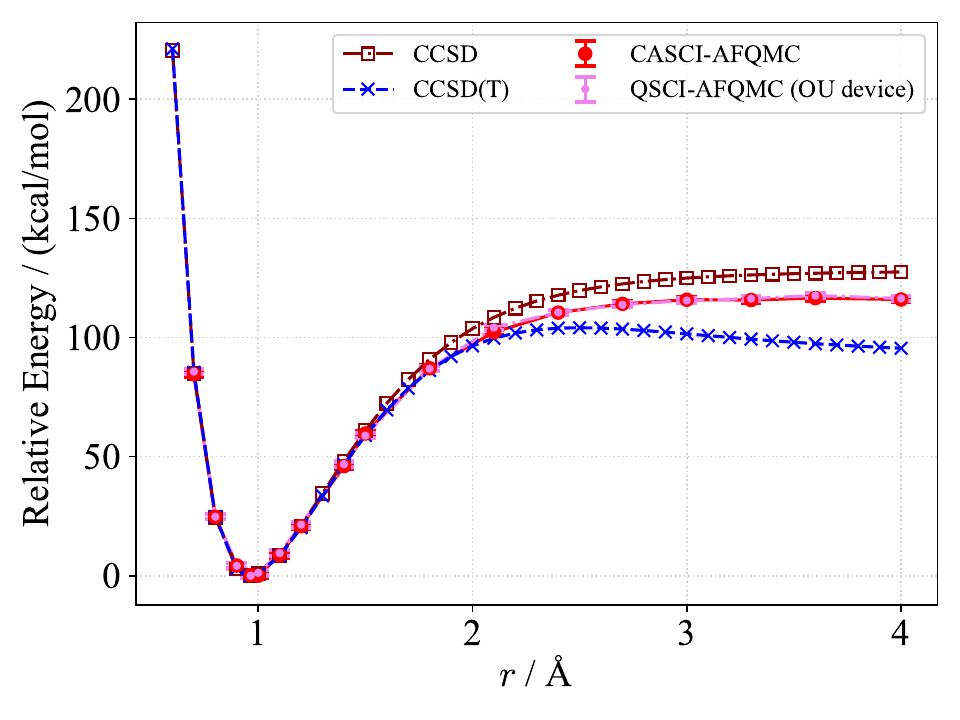}
    \caption{O--H dissociation curve from HEMA at the 6-31++G$**$ basis set level with the active space of (2e, 2o). ``OU Device'' denotes the superconducting quantum computer installed at Osaka University.}
    \label{fig:hema}
\end{figure}

Figure~\ref{fig:hema} displays the relative energies.
Our investigation was performed on the points indicated in the figure, which shows the relative energies whose reference point is the point of the lowest energy for each method.
As with water, the device-based QSCI-AFQMC results correspond closely with those produced by CASCI-AFQMC.
Again, we applied the SCBK mapping and included all sampled bitstrings.
As the bond length increases, CCSD and CCSD(T) deviate from the ph-AFQMC curves, indicating that QSCI-AFQMC and CASCI-AFQMC offer a more accurate depiction of strong correlation effects at larger bond distances.

\subsection{H\textsubscript{4} chain}

\begin{figure}[htbp]
    \centering
    \includegraphics[width=1.0\linewidth, bb=0 0 461 346]{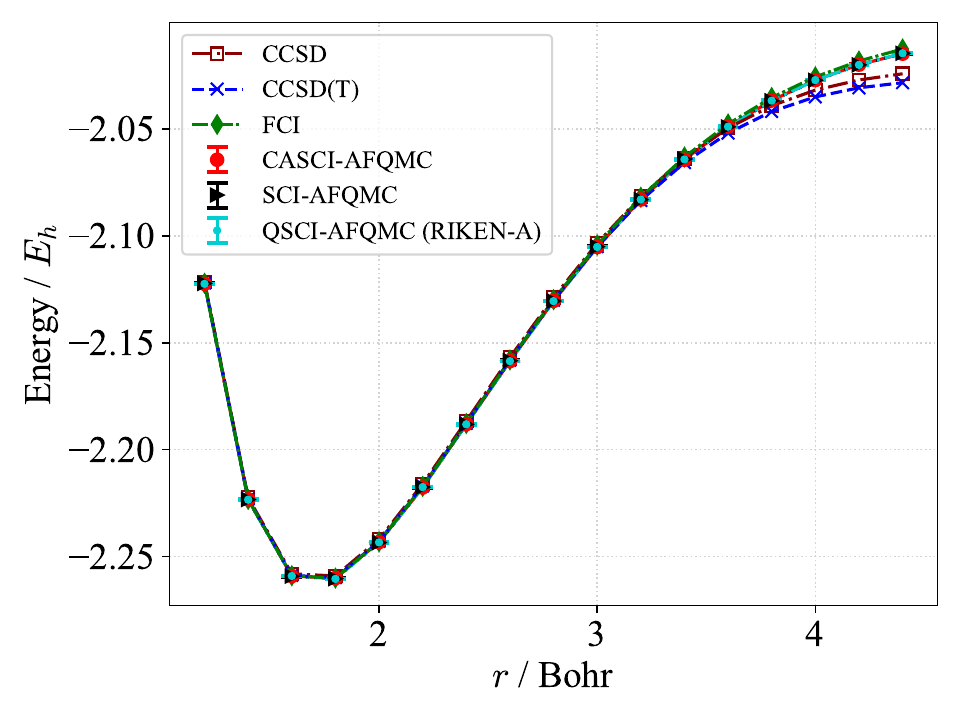}
    \caption{Energy curves for the linear H$_4$ chain at the cc-pVDZ level with an active space of (4e, 4o). The horizontal axis $r$ is the distance between the neighboring hydrogen atoms. ``RIKEN-A'' denotes the superconducting quantum computer at RIKEN.}
    \label{fig:h4}
\end{figure}

Next, we assessed our method on a linear H$_4$ chain, characterized by a larger active space of (4e, 4o). 
The resulting energy curves are shown in Figure~\ref{fig:h4}.
Both the quantum-device-based and simulator-based QSCI-AFQMC calculations accurately track the FCI and CASCI-AFQMC results, whereas CCSD and CCSD(T) start to deviate around $r > 3.5$~Bohr.

\begin{figure}[htbp]
    \centering
    \includegraphics[width=1.0\linewidth, bb=0 0 461 346]{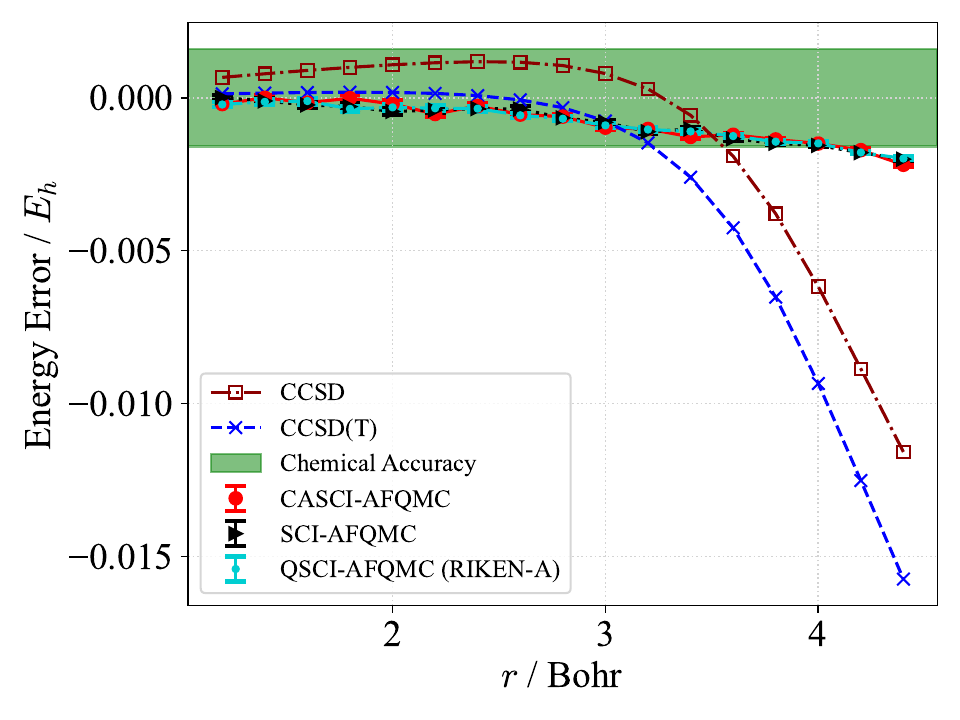}
    \caption{Energy errors relative to FCI for the linear H$_4$ chain. The horizontal axis $r$ is the distance between the neighboring hydrogen atoms. ``RIKEN-A'' denotes the superconducting quantum computer at RIKEN.}
    \label{fig:h4_error}
\end{figure}

For a more detailed comparison, Figure~\ref{fig:h4_error} shows the error relative to FCI.
Although the CCSD and CCSD(T) methods exhibit a decline in accuracy as the distance $r$ increases, the QSCI-AFQMC method maintains chemical accuracy up to $r \leq 4.0$~Bohr and continues to exhibit near-chemical accuracy even at larger distances.
The results obtained from CASCI-AFQMC and SCI-AFQMC were comparable. 
In this system, the configurations selected in both QSCI and SCI are deemed equivalent to the complete active space.

\subsection{N\textsubscript{2}}

Furthermore, we focus on the dissociation of the triple bond in N$_2$, which is recognized as a standard benchmark for multireference (MR) quantum chemical calculations. Our findings are compared against benchmark data published by Lee et al.~\cite{Lee2022twenty,Lee2022Data}, who utilized ph-AFQMC with a CASSCF trial wave function, comparing it with various MR perturbation theories and other advanced methods.

\begin{figure}[ht]
    \centering
    \includegraphics[width=1.0\linewidth, bb=0 0 461 346]{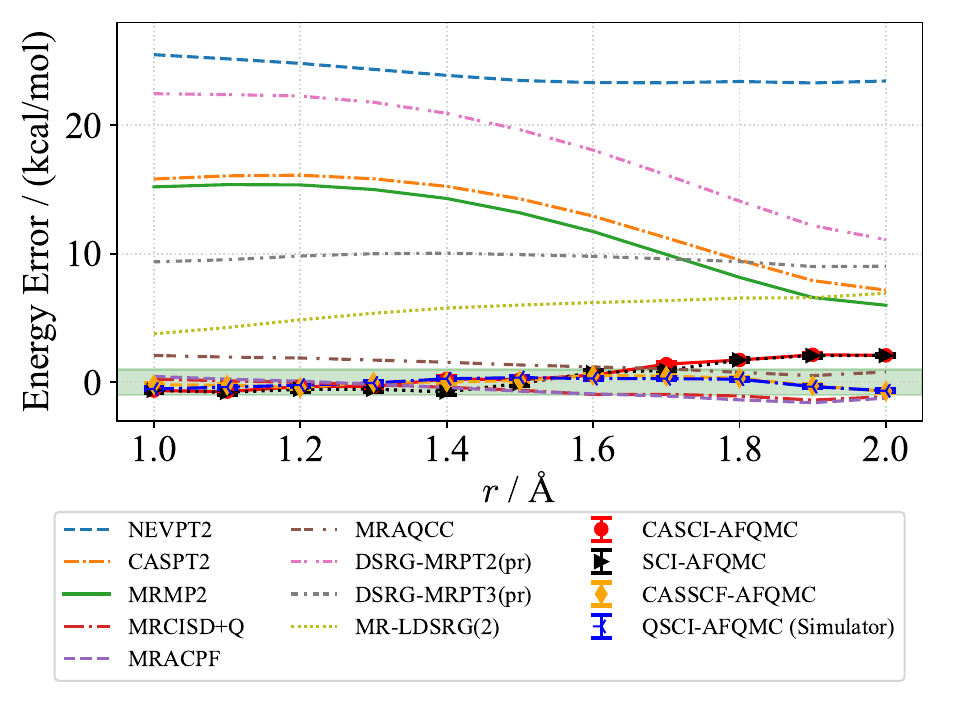}
    \caption{Energy errors for N$_2$ relative to SHCI at the cc-pVTZ level.}
    \label{fig:n2_error}
\end{figure}

Energy deviations from the semistochastic heat-bath configuration interaction (SHCI) for N$_2$ at the cc-pVTZ level are depicted in Figure~\ref{fig:n2_error}.
The present results are compared with the following data from Lee et al.~\cite{Lee2022twenty, Lee2022Data}: SHCI, $N$-electron valence state second-order perturbation theory (NEVPT2), complete active space second-order perturbation theory (CASPT2), multireference M{\o}ller–Plesset second-order perturbation theory (MRMP2), 
multireference configuration interaction with single and double excitations plus the Q correction (MRCISD+Q), multireference averaged coupled-pair functional (MRACPF), multireference average quadratic coupled cluster (MRAQCC), driven-similarity-renormalization-group multireference perturbation theories (DSRG-MRPT2(pr), DSRG-MRPT3(pr)), and linearized multireference driven-similarity-renormalization-group truncated to one- and two-body operators (MR-LDSRG(2)).
As Lee et al. provided data in increments of 0.01~\AA, we performed the comparison at representative bond distances in increments of 0.1~\AA.
In our noiseless simulations, we utilized orbital-optimized variational quantum eigensolver (OO-VQE) calculations~\cite{Mizukami2020orbital}, leveraging a disentangled unitary coupled-cluster singles and doubles ansatz~\cite{kutzelniggQuantumChemistryFock1982,*kutzelniggQuantumChemistryFock1983,*kutzelniggQuantumChemistryFock1985,*bartlettAlternativeCoupledclusterAnsatze1989,*kutzelniggErrorAnalysisImprovements1991,*taubeNewPerspectivesUnitary2006,*evangelista2019exact,Peruzzo2014variational}. Number of sampling shots is set to 10$^6$ for each task.

The OO-VQE-based QSCI-AFQMC faithfully reproduces the results of CASSCF-AFQMC, achieving chemical accuracy compared to SHCI, comparable to the performance of MRCISD+Q.
Significantly, the QSCI-AFQMC calculations outperform MR perturbation theory calculations.
Despite the commendable accuracy of CASCI-AFQMC and SCI-AFQMC, the advantages of orbital optimization become more pronounced at larger bond distances, denoted as $r$, thereby enhancing the trial wave function for ph-AFQMC.

\section{Conclusions} \label{sec:conclusions}

In this study, we advocate for the use of the wave function derived from QSCI as the trial wave function in ph-AFQMC. 
This methodology is referred to as QSCI-AFQMC (more formally QC-QSCI-AFQMC).
Note that QSCI has also been referred to as ``sampling-based quantum diagonalization (SQD)'' in the literature by some research groups since its initial development~\cite{Robledo2024chemistry,Barison2024quantum,Kaliakin2024accurate,Liepuoniute2024quantum,Shajan2024quantum,Kaliakin2025implicit}.
Following this alternative terminology, QSCI-AFQMC could equivalently be denoted as SQD-AFQMC.
Thanks to QSCI, our approach can seamlessly integrate with quantum algorithms (e.g., VQE or quantum phase estimation) that are designed to compute electronic eigenstates on quantum computers.
The subsequent ph-AFQMC calculation performed on a conventional (classical) computer recovered the dynamical electron correlation with high accuracy and parallelization efficiency.
Owing to the QSCI methodology, our method is scalable from the perspective of the quantum resource-efficiency and high flexibility.

We conducted a thorough validation of our methodology across various molecular systems.
For O-H bond dissociation in H$_2$O and the linear dissociation of the H$_4$ chain, the QSCI-AFQMC method produced results that are on par with those obtained from FCI, maintaining chemical accuracy for the majority of data points analyzed.
In the case of N$_2$, the simulations indicated that QSCI-AFQMC consistently achieved a chemical accuracy relative to the SHCI approach and outperformed the conventional MR perturbation theory calculations.

As discussed in Sec.~\ref{sec:results}, the introduction of actual quantum hardware significantly contributes to noise, which directly affects the quality of the sampled bit strings.
Despite these challenges, we successfully demonstrated the practical feasibility of our approach, particularly leveraging SCBK mapping and Cartesian product states to enhance our results.
However, as the number of qubits increases, the exponential rise in possible bit strings could diminish the sampling probability for rare yet crucial configurations.
Therefore, the development of advanced algorithms to improve sampling for these configurations~\cite{Sugisaki2024Hamiltonian,Mikkelsen2024Quantum} will be essential moving forward.

Moreover, recent advancements have verified that ph-AFQMC can be considerably accelerated using graphics processing units (GPUs), thus enhancing its capability to manage molecular systems with a substantial number of orbitals~\cite{Yifei2024GPU}.
Although the current study focused on a system with 222 orbitals, GPU-accelerated ph-AFQMC will facilitate the calculations of even larger orbital spaces.
Consequently, the application of QSCI-AFQMC to large-scale systems stands out as a highly promising avenue for future research.

\section*{Ackowledgements}

We are grateful to Nobuki Inoue and Hanae Tagami for the fruitful discussion.
We also appreciate Satoyuki Tsukano for the technical support.
This work was supported by MEXT Quantum Leap Flagship Program (MEXTQLEAP) Grant No. JPMXS0120319794 and the JST COI-NEXT Program Grant No. JPMJPF2014.
This research was partially facilitated by the JSPS Grants-in-Aid for Scientific Research (KAKENHI) Grant No. JP23H03819.
This paper presents the results obtained using the superconducting quantum computer A installed at RIKEN.
We appreciate the computational resources of the AI Bridging Cloud Infrastructure (ABCI) provided by National Institute of Advanced Industrial Science and Technology (AIST).


\bibliography{apssamp}

\end{document}